\documentclass[twocolumn,aps,amsmath,amssymb,nofootinbib,showpacs]{revtex4}
\usepackage{epsfig}
\usepackage{multirow}
\usepackage{array}
\usepackage{subfigure}
\usepackage{placeins}
\usepackage{pstricks}
\usepackage[nodisplayskipstretch]{setspace}
\usepackage[compact]{titlesec}  
\titlespacing*{\section}{4ex}{4ex}{2ex}
\titlespacing*{\subsection}{4ex}{4ex}{2ex}
\newcommand{\lsim}{\lesssim}
\newcommand{\ord}[1]{\mathcal{O}{(#1)}}

\newcommand{\gsim}{\gtrsim}
\newcommand{\eq}[1]{Eq.~(\ref{#1})}
\newcommand{\beq}{\begin{equation}}
\newcommand{\eeq}{\end{equation}}
\newcommand{\bea}{\begin{eqnarray}}
\newcommand{\eea}{\end{eqnarray}}

\newcommand{\eps}{\varepsilon}

\newcommand{\mzd}{m_{Z_d}}

\newcommand{\diphot}{\gamma\gamma}

\newcolumntype{C}[1]{>{\centering\let\newline\\\arraybackslash\hspace{0pt}}m{#1}}
\newcolumntype{L}[1]{>{\raggedright\let\newline\\\arraybackslash\hspace{0pt}}m{#1}}
\newcolumntype{R}[1]{>{\raggedleft\let\newline\\\arraybackslash\hspace{0pt}}m{#1}}

\def\ptmiss{p\!\!\!\slash_{T}}

\def\gev{\rm GeV}

\def\mev{\rm MeV}

\def\br{{\tt Br}}

\begin{document}

\title{Higgs Decays as a Window into the Dark Sector}
\author{Hooman Davoudiasl$^1$}
\author{Hye-Sung Lee$^{1,2,3}$}
\author{Ian Lewis$^1$}
\author{William J. Marciano$^1$}
\affiliation{$^1$Department of Physics, Brookhaven National Laboratory, Upton, New York 11973, USA\\
$^2$Department of Physics, College of William and Mary, Williamsburg, Virginia 23187, USA\\
$^3$Theory Center, Jefferson Lab, Newport News, Virginia 23606, USA}
\date{\today}
\begin{abstract}
A light vector boson, $Z_d$, associated with a ``dark sector" $U(1)_d$ gauge group has been introduced to explain certain astrophysical observations as well as low energy laboratory anomalies.  In such models, the Higgs boson may decay into $X+Z_d$, where $X=Z,\, Z_d$ or $\gamma$.  Here, we provide estimates of those decay rates as functions of the $Z_d$ coupling through either mass-mixing ({\it e.g.} via an enlarged Higgs mechanism) or through heavy new fermion loops and examine the implied LHC phenomenology.  Our studies focus on the higher $m_{Z_d}$ case, $\gsim$ several GeV, where the rates are potentially measurable at the LHC, for interesting regions of parameter spaces, at a level complementary to low energy experimental searches for the $Z_d$.  We also show how measurement of the 
$Z_d$ polarization (longitudinal versus transverse) can be used to distinguish 
the physics underlying these rare decays.\vspace{0.2in}
\end{abstract}
\pacs{12.60.--i, 14.80.Bn, 14.70.Pw, 13.85.--t\vspace{0.3in}}
\maketitle

%%%%%%%%%%%%%%%%%%%%%%%%%%%%
\section{Introduction}
%%%%%%%%%%%%%%%%%%%%%%%%%%%%
The discovery of a $125-126 ~\gev$ scalar  
state at the Large Hadron Collider (LHC) \cite{:2012gk,:2012gu} appears to have
provided the last missing ingredient, namely the Higgs boson, of
the very successful Standard Model (SM) of particle physics.  
Given its important connection to generation of elementary particle masses, the new 
state, henceforth referred to as the Higgs and
denoted by $H$, is potentially a good place to look for physics beyond the SM.  
For example, $H\to \diphot$ can deviate from  
the SM expectation \cite{Ellis:1975ap,Shifman:1979eb} if new weak scale states 
contribute to the loop processes that mediate the decay.  Early hints of 
such a deviation may already be present.  However, further analysis and more data are necessary to address this question, as the current experimental situation  
does not allow one to draw definite conclusions \cite{Moriond2013}.

In any event, since the SM is insufficient to explain all observations of nature, additional new physics is still required and one can expect more measurements of the Higgs could provide additional surprises.  For example, the presence of dark matter (DM), accounting for more than 80\% of the matter in the Universe~\cite{Beringer:1900zz}, is as-of-yet unexplained.

On general grounds, one may expect that
the DM is part of a larger particle sector whose interactions with the
visible matter (SM) are not completely decoupled.  Such a point of view has
been adopted in explaining some astrophysical data that could
be interpreted as DM signals.
For examples, see Refs.~\cite{Boehm:2003hm,Boehm:2003bt,Boehm:2003ha} 
for an explanation of the 511 keV gamma-rays observed by Integral/SPI \cite{Jean:2003ci} and Ref.~\cite{ArkaniHamed:2008qn} for an explanation of the electron and 
positron excesses observed by ATIC \cite{Chang:2008aa} and PAMELA \cite{Adriani:2008zr}.
Independent of DM considerations, there are also other
possible clues that point to the presence of new physics, in particular the $3.6 \sigma$
discrepancy between the measured value of the muon anomalous 
magnetic moment $g_\mu-2$ and
its predicted value in the SM \cite{Bennett:2006fi,Beringer:1900zz,Czarnecki:2001pv}.  
This and other tentative, less significant hints of beyond the
SM physics may also potentially originate from the same ``dark sector" that
includes DM.

In light of the above considerations, 
we examine here how a $U(1)_d$ gauge interaction 
in the ``dark'' or ``hidden'' sector may manifest itself in Higgs decays.
In particular, we consider the possibility that this Abelian
symmetry is broken, giving rise to a relatively light vector
boson $Z_d$ with mass $\mzd \lsim 10$~GeV.  We further
assume that $Z_d$, being a  hidden sector state, does not couple to 
any of the SM particles including the Higgs directly,
{\it i.e.} SM particles do not carry dark charges.  We do allow for the possibility of
particles in extensions of SM, such as a second Higgs doublet or new heavy leptons, to carry dark charges,
leading to indirect couplings, via $Z-Z_d$ mixing or loop-induced interactions.  
This allows us to assume that the properties of $H$ are close to those of the
SM Higgs, as the current experimental evidence seems to suggest.  For simplicity we will ignore
the possibility of $H$ mixing with other scalars unless specified otherwise.

We will consider decays of the type
\beq
H\to X Z_d \quad (\text{with}~ X = Z,  Z_d, \gamma),
\eeq
and examine the prospects for detecting such
signals at the LHC with emphasis on the $H\to Z Z_d$ decay mode.
Other possibilities will only be briefly commented on.

%%%%%%%%%%%%%%%%%%%%%%%%%%%
\section{Formalism and Phenomenology}
%%%%%%%%%%%%%%%%%%%%%%%%%%%
Quite generally, there are two classes of operators
that contribute to the $H$ decays of interest in our work: 
(A) dimension-3 and (B) dimension-5 operators. 
%\nopagebreak
%%%%%%%%%%%%%%%%%%%%%%%%%%
\subsection {Dimension-3 operators}
\begin{figure}[tb]
\begin{center}
\subfigure[]{
\includegraphics[width=0.23\textwidth,clip]{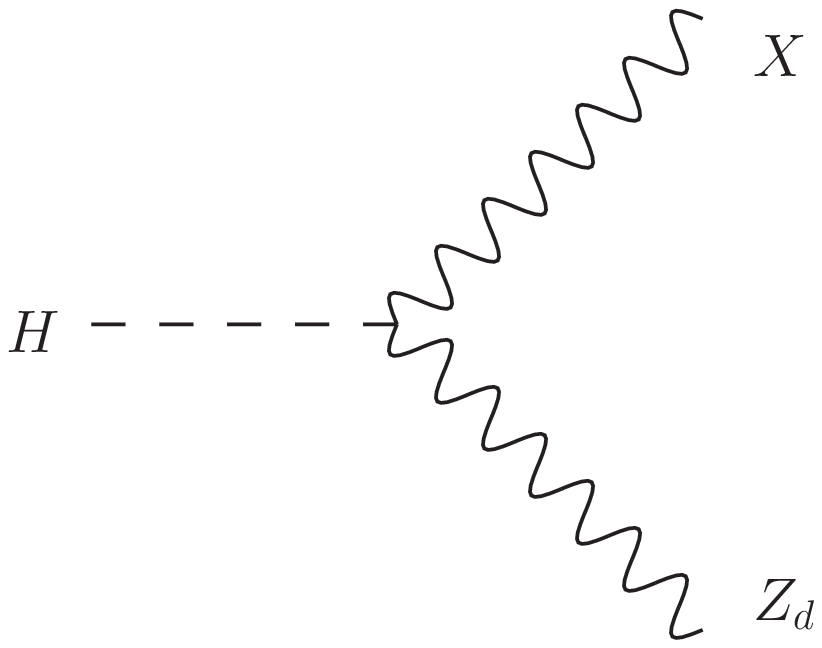}} 
\subfigure[]{
\includegraphics[width=0.23\textwidth,clip]{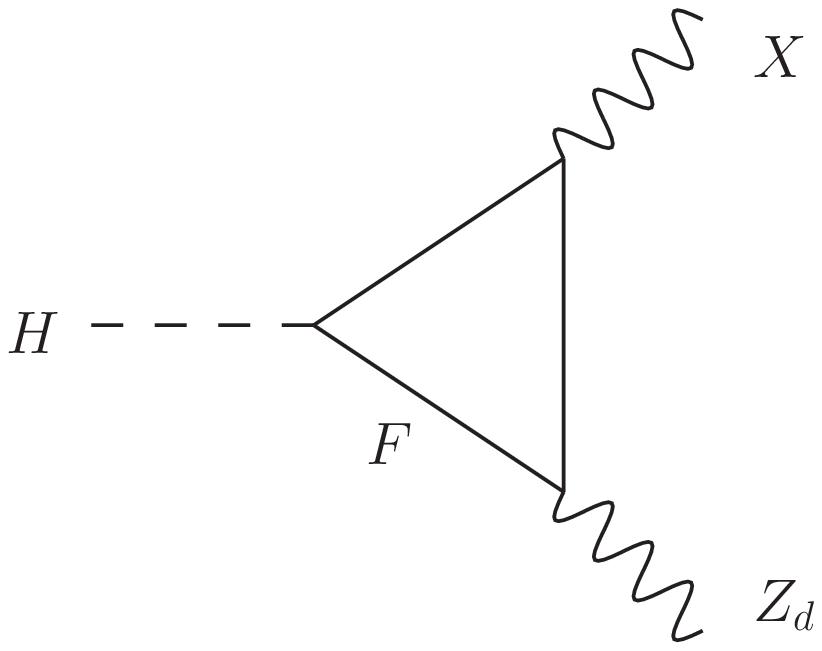}}
\end{center}
\caption{Processes contributing to interactions for (a) case (A) of dimension-3 operator (tree-level mixing) and (b) case (B) of dimension-5 operator (loop-induced decay through a new fermion $F$).}
\label{fig:diagrams}
\end{figure}

%%%%%%%%%%%%%%%%%%%%%%%%%%
We have\footnote{We will not separately list operators that 
involve derivatives of the Higgs \cite{Korchin:2013ifa}, as they can be recast, using equations 
of motion, into one of the forms considered below.}$^,$\footnote{Although not explicitly written, here and henceforth for $X=Z_d$ we assume an additional factor of $1/2$ in our operator definitions to account for an additional combinatoric factor of $2$ in the Feynman rules for identical particles.}

\beq
O_{A,X} = c_{A,X} H X_\mu Z_d^\mu\,,
\label{caseA}
\eeq
with $c_{A,X}$ a coefficient that has mass dimension $+1$.
By gauge invariance $X\neq \gamma$ in this case.  This kind of operator
typically arises from mass mixing between the SM $Z$ and $Z_d$, or by 
Higgs mixing  with another scalar, such as a second Higgs doublet, that carries $U(1)_d$ 
charge and gets a vacuum expectation value (vev).  [See Fig.~\ref{fig:diagrams} (a).]
For a discussion of the latter possibility see, {\it e.g.}, Ref.~\cite{Gopalakrishna:2008dv} where Higgs mixing with a SM singlet scalar charged under $U(1)_d$ has been considered.
Whatever its origin, for processes mediated by the operator $O_{A,X}$, the boosted $Z_d$ final states will be primarily longitudinal.

When a mass mixing is involved, it is straightforward to understand the manifest longitudinal polarization.
The longitudinal polarization of the physical eigenstate $Z_d$ inherits a component of a SM Nambu-Goldstone boson (NGB) from the SM $Z$.  Hence, for a high energy longitudinally polarized $Z_d$ we can make the replacement in Eq.~(\ref{caseA}):
\begin{equation}
Z_d^\mu \rightarrow \partial^\mu\phi/m_{Z_d}+\ord{m_{Z_d}/E_{Z_d}},
\label{goldstone}
\end{equation}
where $\phi$ is a NGB and $E_{Z_d}$ is the energy of the $Z_d$.  Hence, when the $Z_d$ is highly boosted the longitudinal mode is enhanced over the transverse polarizations by the energy dependence of the derivative coupling of the NGB.

For operators of type $O_{A,X}$ in \eq{caseA}, we will focus on $X=Z$.  Such interactions
are typically associated with mixing.  For example, the mass term for 
$Z$-$Z_d$ mixing can be parametrized as $\eps_Z \, m_Z^2 Z Z_d$, with   
\beq
\eps_Z = \frac{\mzd}{m_Z} \delta \,,
\label{epsZ}
\eeq
where $\delta$ is a model dependent parameter.  (We will assume
that $Z_d$ is light compared to $Z$ in this work.)  The vector boson  
$Z_d$ couples to the weak neutral current, like the SM $Z$ boson, 
with a coupling suppressed by $\eps_Z$.  This coupling can then provide a new 
source of parity violation \cite{Davoudiasl:2012ag} and couplings of $Z_d$ to 
non-conserved currents.  At energies $E\gsim m_{Z_d}$, the longitudinal $Z_d$
has enhanced couplings of order $(E/\mzd) \eps_Z$ which dominate its 
weak scale processes, as implied by the Goldstone boson equivalence
theorem \cite{GET}.  Therefore, we expect the Higgs decay
$H\to Z Z_d$ mediated by $O_{A,X}$ to be dominated by longitudinal
vector bosons.  As discussed in Refs.~\cite{Davoudiasl:2012ag,Davoudiasl:2012qa}, 
this scenario can be constrained by various 
experiments, both at low and high energies.  (See Ref.~\cite{Kumar:2013yoa} 
for a recent review of the relevant low energy parity violation experiments.)  

%%%%%%%%%%%%% FIGURE %%%%%%%%%%%%%
%%%%%%%%%%%%%%%%%%%%%%%%%%%%%%%%%%

Roughly speaking, rare $B$ and $K$ decays suggest $\delta^2 \lsim 10^{-5}$ for $m_{Z_d} \ll 5 ~\gev$ (a scale set by the $B$ meson mass) while the good agreement  between precision $Z$ pole property measurements and SM predictions imply $\delta^2 < \text{few}~ \times 10^{-4}$ for all $m_{Z_d}$ \cite{Davoudiasl:2012ag}.
For comparison, the good agreement  between early $H$ decay data at the LHC already suggests $\delta^2 \lsim 10^{-4}$ for some values of $m_{Z_d}$ \cite{Davoudiasl:2012ag} which is competitive with precision $Z$ pole studies and the (somewhat model dependent) rare meson decays.
Hence, continued future searches for $H \to X Z_d$ hold the promise of providing a sensitive, 
unique probe of $Z-Z_d$ mixing, particularly in the higher $m_{Z_d}$ region above a few $\gev$ where rare flavor decays are not applicable.

We briefly compare the rare Higgs branching ratios for the typical case of $Z$-$Z_d$ mass mixing.
As studied in Ref.~\cite{Davoudiasl:2012ag}, we have $\br (H \to Z Z_d) \simeq 16 \delta^2$. For the given bound on $\delta^2$, this branching ratio is possibly comparable to $\br (H \to \gamma\gamma) \simeq 2.3 \times 10^{-3}$ for the SM Higgs of $125 ~\gev$. The decay
$H \to Z_d Z_d$ is negligible, as it is doubly suppressed by $\delta^2$ ($\Gamma (H \to Z_d Z_d) / \Gamma (H \to Z Z_d) \simeq 5 \delta^2$).  Finally, there is no $H \to \gamma Z_d$ at leading order.

%%%%%%%%%%%%%%%%%%%%%%%%
\subsection{Dimension-5 operators} 
%%%%%%%%%%%%%%%%%%%%%%%%
%%%%%%%%%%%%% FIGURE %%%%%%%%%%%%%
\begin{figure}[b]
\begin{center}
\includegraphics[width=0.3\textwidth,clip]{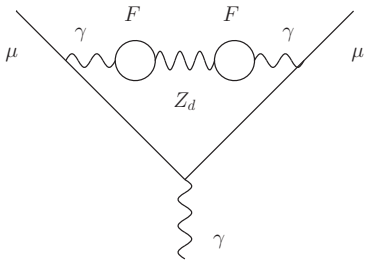}
\end{center}
\caption{$Z_d$ contribution to the $g_\mu - 2$. The new fermion ($F$) loop-induced $\gamma$-$Z_d$ mixing is explicitly shown.}
\label{fig:g-2}
\end{figure}
%%%%%%%%%%%%%%%%%%%%%%%%%%%%%%%%%%

These operators can be CP even or odd.  The CP even interaction has the form  
\beq
O_{B,X}=c_{B,X} H X_{\mu\nu}Z_d^{\mu\nu}\,,
\label{caseB}
\eeq
where $X_{\mu\nu} = \partial_\mu X_\nu-\partial_\nu X_\mu$ is the field strength associated 
with $X$.  The CP odd interaction can be written as 
\beq
\tilde{O}_{B,X}=\frac{\tilde{c}_{B,X}}{2} 
\eps_{\mu\nu\rho\sigma}H X^{\mu\nu}{Z_d}^{\rho \sigma}\,.
\label{caseBtil}
\eeq
Here, $c_{B,X}$ and $\tilde{c}_{B,X}$ have mass dimension $-1$ 
and are model dependent.
Generically, these interactions result from 
1-loop processes [see Fig.~\ref{fig:diagrams} (b)].  For example, in models with vector-like
fermions carrying $SU(2) \times U(1)_Y \times U(1)_d$ quantum
numbers such as the heavy lepton model of Ref.~\cite{Davoudiasl:2012ig}, 
these operators can naturally occur.  
The CP odd interaction in \eq{caseBtil} arises if the new 
fermions have complex couplings to the Higgs 
that introduce CP violating physical phases \cite{Voloshin:2012tv}.   

%%%%%%%%%%%%%%%%%%%%%%%%%%%%%%%%%%
%%%%%%%%%%%%%%%%%%%%% FIGURE %%%%%%%%%%%%%
\begin{figure*}[tb]
\begin{center}
\subfigure[]{
      \includegraphics[height=0.45\textwidth,clip]{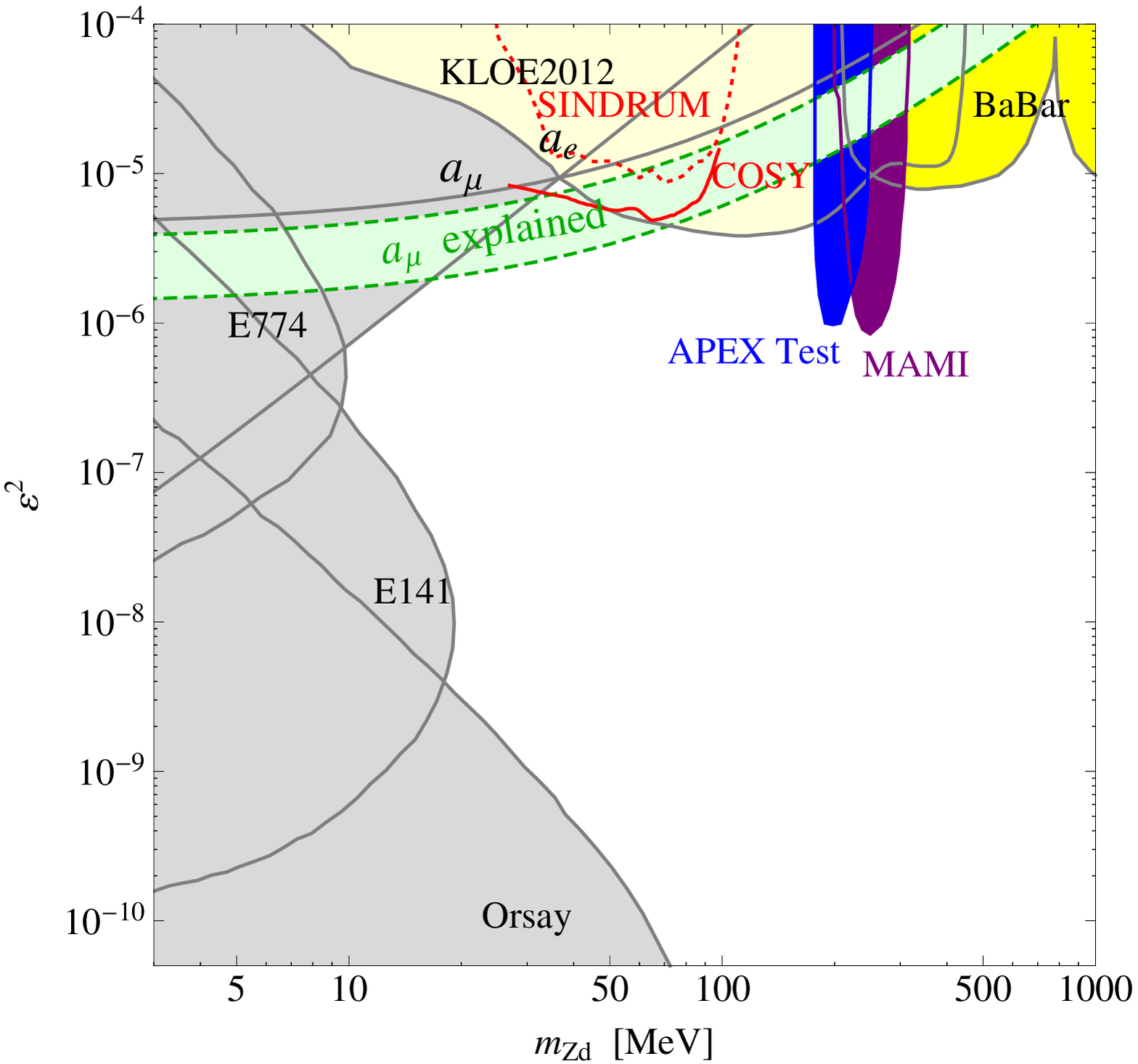}
} ~
\subfigure[]{
      \includegraphics[height=0.45\textwidth,clip]{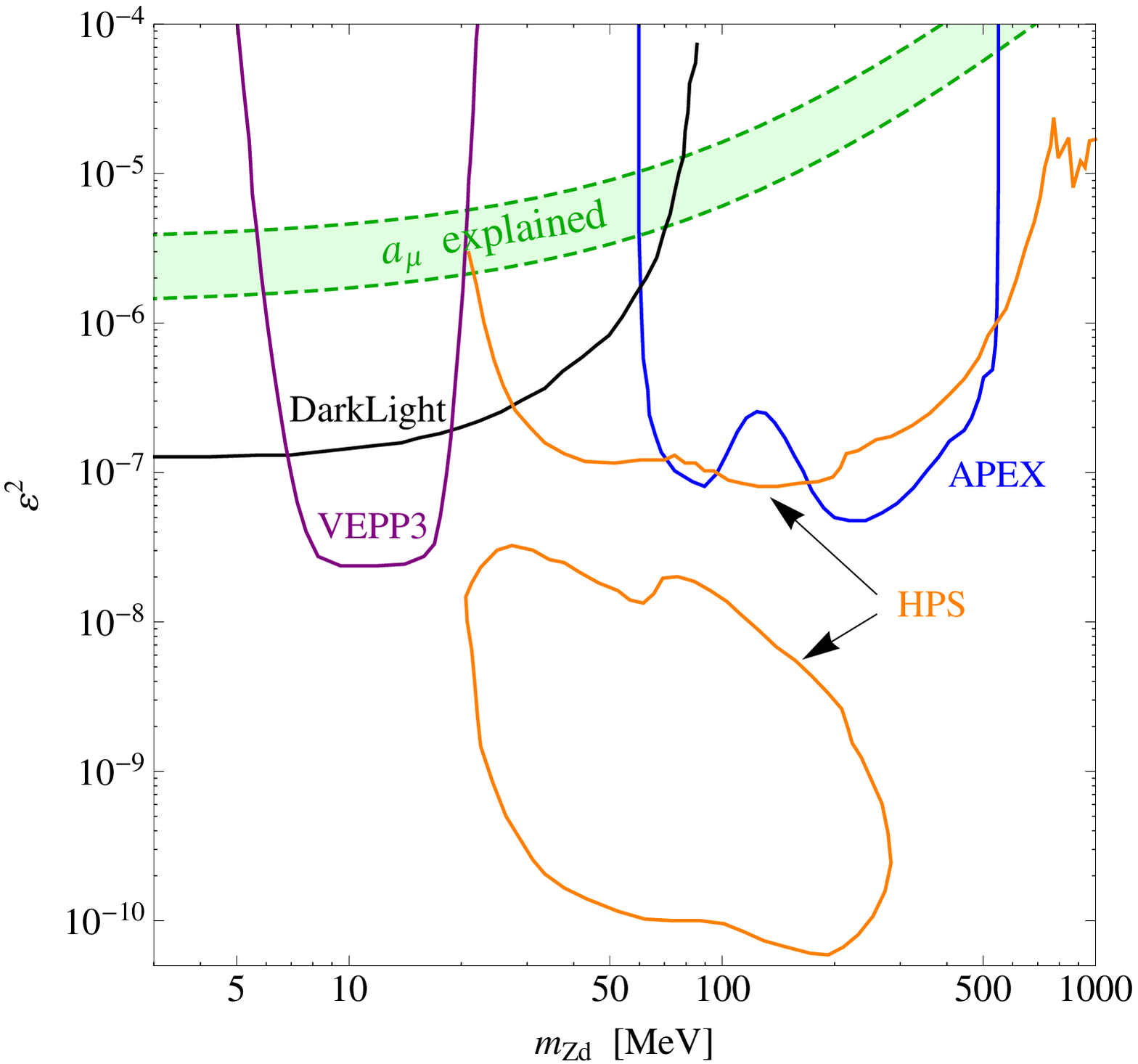}
}
\caption{(a) Shaded regions of $\eps^2 - m_{Z_d}$ parameter space ruled out by various experimental constraints. The green band shows the parameter region (90\% CL) that can explain the $3.6 \sigma$ discrepancy in $g_\mu - 2$ \cite{Davoudiasl:2012ig}. (b) The same parameter space showing the sensitivity of future direct light $Z_d$ gauge boson searches. Such search constraints generally assume $\br (Z_d \to \ell^+ \ell^-) \simeq 1$. In the case of VEPP3 \cite{Wojtsekhowski:2012zq}, the anticipated sensitivity is independent of the $\br (Z_d \to \ell^+ \ell^-)$.}
\label{fig:parameterSpace}
\end{center}
\end{figure*}
%%%%%%%%%%%%%%%%%%%%%%%%%%%%%%%%%%

One-loop effects from the aforementioned vector fermions typically result in   
kinetic mixing between $U(1)_d$ and $U(1)_Y$ \cite{Holdom:1985ag,darkphoton,nonabelian}, 
parametrized by
\beq
\frac{\eps}{2 \cos \theta_W} B_{\mu\nu}Z_d^{\mu\nu}\,,
\label{kinmix}
\eeq
where $\theta_W$ is the weak mixing angle and $B_{\mu\nu}$ is the 
hypercharge field strength tensor.  
For a loop-generated $\eps$, we would generally 
expect $\eps \sim e g_d /(16\pi^2) \sim \ord{10^{-3}}$ assuming unit charges; $e$ is the 
electromagnetic coupling constant and $g_d$ is 
the $U(1)_d$ gauge coupling.  
The vector $Z_d$ can couple to the electromagnetic current like a photon with a coupling suppressed by $\eps$.

A light $Z_d$ with small coupling to the electromagnetic current can contribute to the $g_\mu - 2$.
The relevant diagram, explicitly showing the $\gamma$-$Z_d$ mixing through heavy fermion ($F$) loops, is given in Fig.~\ref{fig:g-2}.
A similar loop-induced vertex amplitude is illustrated in Fig.~\ref{fig:diagrams} (b) for the Higgs decay into $Z_d$.

The $Z_d$ contribution to $g_\mu-2$ can explain the current $3.6 \sigma$ deviation of the muon anomalous magnetic moment from the SM prediction \cite{Fayet:2007ua,Pospelov:2008zw} in a parameter region indicated by the green band in the $\eps^2 - m_{Z_d}$ parameter space in Fig.~\ref{fig:parameterSpace} (a).
As the figure shows, roughly $\eps^2 \sim 10^{-6} - 10^{-5}$ and $m_{Z_d} \sim 20 - 50 ~\mev$ are left, after various experimental constraints with the most recently updated values are imposed: electron $g-2$ \cite{Davoudiasl:2012ig,Endo:2012hp}, beam dump experiments \cite{Bjorken:2009mm,Andreas:2012mt}, meson decays into $Z_d$ ($\Upsilon$ decays at BaBar \cite{Bjorken:2009mm}, $\phi$ decays at KLOE \cite{Babusci:2012cr}, $\pi^0$ decays at SINDRUM \cite{MeijerDrees:1992kd,Gninenko:2013sr} and WASA-at-COSY \cite{Adlarson:2013eza}), and fixed target experiments (MAMI \cite{Merkel:2011ze}, APEX \cite{Abrahamyan:2011gv}).
The requisite value of $\eps$ for the $g_\mu - 2$ explanation can naturally arise from 1-loop diagrams if $g_d \approx e$.
Figure \ref{fig:parameterSpace} (b) shows the expected sensitivities of future $Z_d$ searches beyond existing bounds in the same parameter space \cite{McKeown:2011yj}.

As we discussed, the low mass region ($m_{Z_d} \lsim 1 ~\gev$) is particularly well motivated by the $g_\mu - 2$ discrepancy but it is being thoroughly explored by direct production at electron facilities (such as the one at Jefferson Lab in the US, and the one at Mainz in Germany), as well as rare meson decays.
In this paper, we will concentrate on the higher mass region ($m_{Z_d} \gsim 1 ~\gev$) which is potentially accessible in Higgs decays.

Assuming that the couplings in Eqs.~(\ref{caseB}) and (\ref{caseBtil}) are induced 
by loops of vector-like fermions, with $SU(2)\times U(1)_Y$ preserving masses $m_F$, 
we expect $c_{B,X}$, ${\tilde c}_{B,X} \to 0$ as $m_F \to \infty$.  
The masses of these fermions are shifted after electroweak symmetry 
breaking, due to their Yukawa interactions with the Higgs.  For 
$m_F \sim {\rm few}\times 100$~GeV and Yukawa couplings $y_F\sim 1$ the lightest new fermions 
can have masses of $\ord{m_Z}$~\footnote{For unstable fermions with $m_H > m_F\sim \ord{m_Z}$, the Higgs decay channel $H\rightarrow F F^*$ opens up.  Since $m_F+m_{W,Z}>m_H$, the full decay chain proceed through an off-shell $W$ or $Z$: $H\rightarrow F F^*\rightarrow F F_0 W^*/Z^*\rightarrow F F_0 f' \bar{f}$, where $f,f'$ are SM fermions and $F_0$ is either a SM lepton or a new fermion stable on collider timescales.  In either case, this is a four-body Higgs decay and is suppressed relative to the dominant SM decay channels.  If $F$ is stable, then the decay $H\to F F^*$ is unavailable for this mass.} and [for $\ord{1}$ phases] we get   
\beq
|c_{B,X}| \sim |{\tilde c}_{B,X}| \sim \frac{g_w g_d y_F}{16 \pi^2 \, m_Z}\,,
\label{cB}
\eeq
with $g_w$ a typical electroweak coupling constant.  
Such vector-like fermions [but not necessarily charged under $U(1)_d$]  
have been recently proposed as a natural solution to the larger than expected diphoton 
rate in the early Higgs data at the LHC (for some examples, see Refs.~\cite{Dawson:2012di,
Carena:2012xa,Bonne:2012im,An:2012vp,Joglekar:2012vc,ArkaniHamed:2012kq,
Almeida:2012bq,Kearney:2012zi,Ajaib:2012eb,Dawson:2012mk,Davoudiasl:2012tu}).

As discussed in Ref.~\cite{Davoudiasl:2012ig}, with the additional assumption 
of being charged under $U(1)_d$, the aforementioned fermions 
are also well motivated as mediators of the $U(1)_Y$ and $U(1)_d$ kinetic mixing.  
In this framework, we can estimate 
the $H \to \gamma Z_d$, $H\to Z Z_d$, and $H \to Z_d Z_d$ rates.
For $g_d \approx e$, as motivated 
by the above discussion of $g_\mu - 2$, and assuming roughly equal values for $c_{B,X}$ and $\tilde{c}_{B,X}$
it is estimated that $0.1 \, \br (H \to \gamma \gamma)\approx \br(H\to\gamma Z_d)\approx 2\, \br(H\to Z_d Z_d)\approx 10\, \br(H\to Z Z_d)$ if the vector-like fermions are to explain the Higgs to diphoton decay rate 
at about 1.5 times the SM prediction\footnote{At leading order, 
$W$ loops do not contribute to $O_{B,X}$ and $\tilde{O}_{B,X}$ 
while their contribution is dominant over the fermion 
loop contributions in the SM $H \to \diphot,\, Z \gamma$ 
processes.  The $H\to XZ_d$ rates would then likely be comparatively suppressed.}. The suppression of $\br(H\to Z Z_d)$ relative to $\br(H\to Z_d Z_d)$ is due to an additional phase space suppression of having a massive final state particle, and the factor of two between $\br(H\to\gamma Z_d)$ and $\br(H\to Z_d Z_d)$ is due to a symmetry factor from identical final states.

Given the preceding discussion, for $\br (Z_d \to \ell^+\ell^- ) > \br (Z \to \ell^+\ell^- )$, 
one could expect signals of $H\to X Z_d$ from the
interactions in Eqs. (\ref{caseB}) and (\ref{caseBtil}) at the LHC, 
with current or near future levels of statistics. 
The $Z_d$ can mimic the promptly converted photon ($\gamma \to e^+ e^-$) 
although the small nonzero mass and a production 
vertex near the beam can be used to distinguish the new $Z_d$ events.

  We also note that 
the final state $Z_d$ from decays mediated by $O_{B,X}$ and $\tilde{O}_{B,X}$ 
would be primarily transversely polarized. This can be understood by making the replacement of the longitudinally polarized $Z_d$ in Eq.~(\ref{goldstone}) into operators $O_{B,X}$ and $\tilde{O}_{B,X}$.  In this case, the longitudinal polarizations completely decouple up to $\ord{m_{Z_d}/E_{Z_d}}$.  Hence, for a highly boosted $Z_d$, the transversely polarized $Z_d$ dominates the interactions mediated by $O_{B,X}$ and $\tilde{O}_{B,X}$.

%%%%%%%%%%%%%%%%%%%%%%%%%%%%
\section{Parameterization and Setup}
%%%%%%%%%%%%%%%%%%%%%%%%%%%%
Models that could give rise to operators $O_{A,X}$, $O_{B,X}$, and 
$\tilde{O}_{B,X}$ have recently been studied, for example, in
Refs.~\cite{Davoudiasl:2012ag,Davoudiasl:2012ig,Voloshin:2012tv,Lee:2013fda}.  Here, we briefly 
explain our parameterization and set up our notation.
The general decay widths for each case can be found in our Appendix.

{\bf Class (A)}:  Regardless of the origin of $O_{A,X}$ 
(be it mixing between scalars or vectors), following $Z - Z_d$ mixing we will parameterize 
the interaction in \eq{caseA} by
\beq
c_{A,X} = \frac{g}{\cos \theta_W}\eps_Z\, m_Z\,,
\label{kappaA}
\eeq
where $g$ is the $SU(2)_L$ coupling constant.

{\bf Class (B)}:  
In analogy with $c_{A,X}$,
we parameterize the strength of interactions from $O_{B,X}$ and ${\tilde O}_B$ by
\beq
c_{B,X}=-\frac{g}{2\cos \theta_W}(\kappa_X/m_Z) 
\label{kappaB}
\eeq
and
\beq
{\tilde c}_{B,X}=\frac{g}{2\cos \theta_W}({\tilde \kappa}_X/m_Z) \,,
\label{kappaBtilde}
\eeq
respectively, where $\kappa_X$ and ${\tilde \kappa_X}$ are dimensionless model dependent constants.  The normalizations 
have been chosen for ease of notation in our later results.  

As previously pointed out, the interactions mediated by class (A) and (B) operators result in
final states that are dominated by the longitudinal and transverse polarizations of the vector bosons, respectively.  
{\it This effect can be used as a diagnostic probe of the underlying microscopic process to determine whether
they result from class (A) or class (B).}  The operators 
$O_{B,X}$ and $\tilde O_{B,X}$ can be further disentangled 
if the angular distribution of the final state particles are considered.
For the Higgs to diphoton decay, for example, see Ref.~\cite{Voloshin:2012tv}.

The decay products of a very light $Z_d$ in the decay chain $H\rightarrow Z Z_d\rightarrow 4\ell$ are highly collimated and will be difficult to isolate.  Also, the decay $H\rightarrow Z Z_d\rightarrow 4\ell$ can be expected to have already appeared in the $H\rightarrow ZZ^*\rightarrow 4\ell$ signal, where the $Z_d$ would appear as a peak in the invariant mass distribution of the $Z^*$.  However, ATLAS~\cite{ATLAS:2013nma} and CMS~\cite{CMS:xwa} both require that $m_{Z^*}>12$~GeV in their $H\rightarrow 4\ell$ analysis. Hence, the mass range we consider in our LHC study is primarily $m_{Z_d} \approx 5 - 10 ~\gev$ 
which is  complementary to searches of lighter $Z_d$ in low energy experiments; 
see Fig.~\ref{fig:parameterSpace}.   For $Z_d$s with mass below $5$~GeV, the leptons from $Z_d$ decay will be highly collimated.   
Our examination of Higgs decays to light vector bosons at the LHC is also complementary to 
studies based on heavy (TeV scale) $Z'$ scenarios 
(for example, see Refs.~\cite{Agashe:2007ki,Barger:2009xg}).  Before closing this section, 
we would like to point out that the size of $\br(Z_d\to \ell^+ \ell^-)$, $\ell=e,\mu$, is model dependent.  For example,
if kinetic mixing between $U(1)_d$ and $U(1)_Y$ ($\varepsilon \neq 0$) dominates, one might expect $\br(Z_d\to \ell^+ \ell^-) \simeq 0.3$ for the $5-10$~GeV mass range while for $Z - Z_d$ mixing dominance it could be smaller \cite{Davoudiasl:2012ag}.
Because of that arbitrariness, we will give our results in terms of $\delta^2 \times \br(Z_d \to \ell^+ \ell^-)$.

%%%%%%%%%%%%%%%%%%%%%%%%%%%%
\section{Numerical Analysis}
%%%%%%%%%%%%%%%%%%%%%%%%%%%%
%%%%%%%%%%%%% FIGURE %%%%%%%%%%%%%
\begin{figure}[tb]
\begin{center}
\includegraphics[width=0.45\textwidth,clip]{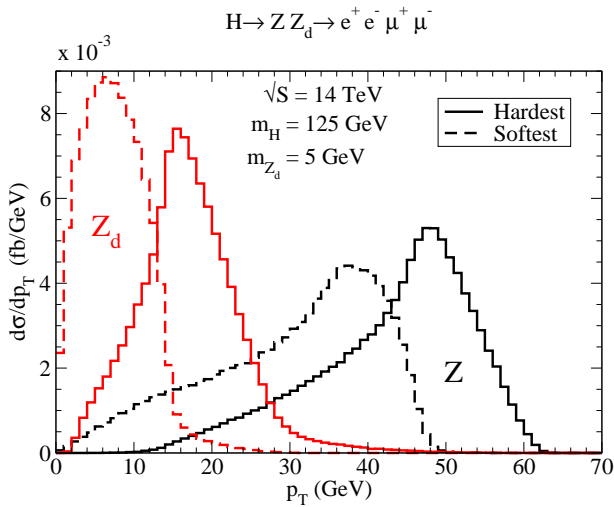}
\end{center}
\caption{Transverse momentum distribution of the hardest (solid) and softest (dashed) leptons identified as originating from the $Z$ (black) and $Z_d$ (red/grey) for the $e^+e^-\mu^+\mu^-$ final state.  We use the parameter choices of Eqs.~(\ref{masses}, \ref{params}).  No energy smearing or cuts have been applied.}
\label{fig:pT}
\end{figure}
%%%%%%%%%%%%%%%%%%%%%%%%%%%%%%%%%%

We now present the results of our collider simulation.  We study the observability of
\begin{equation}
pp\rightarrow H \rightarrow Z Z_d \rightarrow \ell^+_1\ell^-_1 \ell^+_2\ell^-_2,
\end{equation}
at the $\sqrt{s}=14$~TeV LHC, where $\ell_1$ and $\ell_2$ are electrons or muons.

The decay $H \to Z Z_d$ is an interesting channel to study as it can arise in both 
class (A) and class (B) type interactions.  The $H \to \gamma Z_d$ mode is not relevant for class (A), 
since gauge invariance does not allow it.
While the $H \to Z_d Z_d$ decay may also occur for both classes of interactions, 
it will be doubly suppressed if it results only from $Z$-$Z_d$ mass mixing.  
The decay widths for all three modes are presented in the Appendix, using the 
parametrization of the previous section ($\eps_Z$, $\kappa_X$, $\tilde \kappa_X$).

The region of interest is a relatively light $Z_d$, and so, unless otherwise noted, we use a $Z_d$ and Higgs mass of 
\begin{eqnarray}
m_{Z_d}\,=\,5\,{\rm GeV}\quad{\rm and}\quad m_H\,=\,125~{\rm GeV}.
\label{masses}
\end{eqnarray}
In our analysis we find that the kinematic cuts affect events from operators of class (A) and (B) somewhat differently.  For simplicity, the results of the numerical simulation are presented in detail with the coupling constants
\begin{eqnarray}
\delta^2\times \br(Z_d \to \ell^+ \ell^-)=10^{-5} \quad {\rm and}\quad\kappa_Z=\tilde{\kappa}_Z=0.\label{params}
\end{eqnarray}
More general cases can be obtained by rescaling the overall coupling and taking into account how the kinematic cuts effect the two classes of events, as is detailed at the end of our analysis.

Using the results of Ref.~\cite{Davoudiasl:2012ag}, the above choice corresponds to 
$\br(H\to Z Z_d) \approx 2\times 10^{-4}/\br(Z_d \to \ell^+ \ell^-)$.  
As mentioned before, when kinetic mixing dominates, 
the branching ratio of $Z_d$ into leptons is typically given by 
$\br(Z_d \rightarrow e^+e^-) \simeq \br(Z_d \rightarrow \mu^+\mu^-) \simeq 0.15$.

%%%%%%%%%%%%% FIGURE %%%%%%%%%%%%%
\begin{figure*}[tb]
\begin{center}
\includegraphics[width=0.5\textwidth,clip]{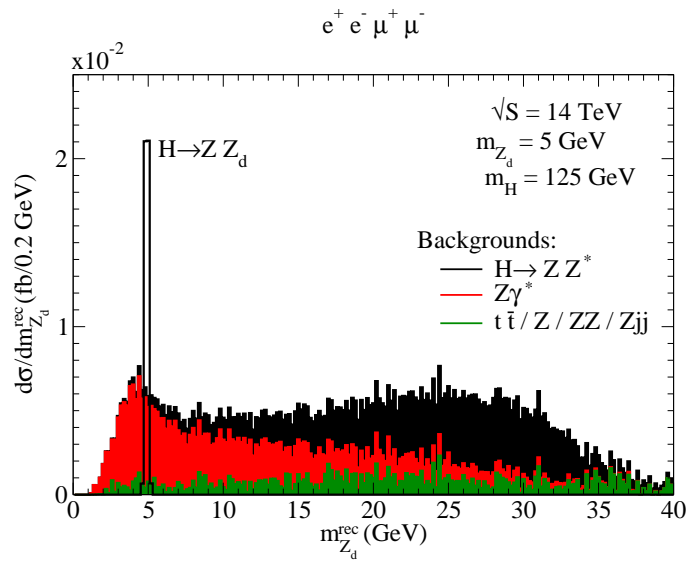}
\end{center}
\caption{Distribution of the reconstructed $Z_d$ invariant mass, $m_{Z_d}^{rec}$, for both major background (shown cumulatively) and signal for the $e^+e^-\mu^+\mu^-$ final state.  The parameters of Eqs.~(\ref{masses}, \ref{params}) were used.  Energy smearing, triggers, and cuts in Eqs.~(\ref{acc.EQ} - \ref{Zmass.EQ}) were applied.}
\label{fig:dileptons}
\end{figure*}
%%%%%%%%%%%%%%%%%%%%%%%%%%%%%%%%%%

We now discuss how to isolate our signal from backgrounds.  All signal and backgrounds were simulated using MadGraph 5~\cite{Alwall:2011uj} with the CTEQ6L parton distribution set~\cite{Pumplin:2002vw}.  The operators in Eqs.~(\ref{caseA},\ref{caseB},\ref{caseBtil}) with the parameterization in Eqs.~(\ref{kappaA} - \ref{kappaBtilde}) were implemented into MadGraph 5 using FeynRules~\cite{Christensen:2008py}.

To isolate the signal, it will be useful to implement full event reconstruction.  Hence, we need to identify which opposite sign, same flavor lepton pair originates from the $Z$ and which originates from the $Z_d$.  Since both the $Z$ and $Z_d$ have narrow widths, we expect one pair of leptons to reconstruct the $Z$ mass and the other pair the $Z_d$ mass.  Since the $Z_d$ mass is not known {\it a priori}, the $Z$ boson mass is utilized to identify the decay products.  We calculate the invariant mass of all opposite sign, same flavor lepton pairs and identify the pair that closest reconstructs the $Z$ pole as originating from the $Z$ boson.  The remaining pair is identified as originating from the $Z_d$.  We will see that the invariant mass distribution of the lepton pair identified as originating from the $Z_d$ will permit a measurement of $m_{Z_d}$.

Detector resolution effects are simulated via Gaussian energy smearing applied to leptons and jets with a standard deviation parameterized as:
\begin{eqnarray}
\frac{\sigma(E)}{E}=\frac{a}{\sqrt{E}}\oplus b,
\end{eqnarray}
where $\sigma(E)$ is the energy resolution at energy $E$, $\oplus$ represents addition in quadrature, and all energies are measured in GeV.  For leptons (jets) we take the ATLAS values~\cite{Voss:2009zz} $a=10\%$ ($50\%$) and $b=0.7\%$ ($3\%$).  The energy resolution of electrons and muons have different dependencies on electromagnetic (EM) calorimetry and charged particle tracking.  However, we use the uniform values for EM calorimetry energy resolution for all final state leptons, which, for the energies under consideration, is more conservative than tracking capabilities.

For events to be triggered on, they must pass minimum acceptance cuts on the rapidities, $\eta$, and transverse momenta, $p_T$, of the final state particles.  Typically, for leptons a $p_T$ cut around $20$~GeV is used.  However, for our signal, two of the leptons originate from a low mass resonance and are expected to have low transverse momenta. In Fig.~\ref{fig:pT}, we show the transverse momentum distributions of the hardest (solid) and softest (dashed) leptons identified as originating from the $Z$ (black) and $Z_d$ (red/gray) for the $e^+e^-\mu^+\mu^-$ final state.  Distributions for the $2e^+2e^-$ and $2\mu^+2\mu^-$ final states are similar.  The leptons from the $Z$ decay are typically harder than those from the $Z_d$.  This can be understood by noting that the momentum of the $Z$ and $Z_d$ in the Higgs rest frame is $|p|\approx30$~GeV.  Hence, the energy of the $Z$ is dominated by the $Z$ mass, $m_Z$, while the energy of a light $Z_d$ is dominated by $|p|$.  The decay products of the $Z$ then have higher transverse momentum than the decay products of the $Z_d$ and typically peak near $m_Z/2$.  
Based on these considerations, we apply transverse momentum and rapidity cuts on all final state leptons:
\begin{equation}
p_T^{\ell}\,>\,4\,{\rm GeV}\quad{\rm and}\quad|\eta^\ell|\,<\,2.5.
\label{acc.EQ}
\end{equation}
To trigger on an event, ATLAS and CMS typically require that at least one final state particle have $p_T$ 
larger than that required by Eq.~(\ref{acc.EQ}). We follow the ATLAS triggers~\cite{:2012gk} on $H\rightarrow Z Z^*\rightarrow 2\ell^+ 2\ell^-$ events and require that one lepton passes a transverse momentum threshold of $24$~GeV or two leptons have a minimum transverse momentum of $13$~GeV each.

%%%%%%%%%%%%% TABLE %%%%%%%%%%%%%
\begin{table*}[htb]
\centering
\caption{Signal and background cross sections (in fb) with consecutive cuts, and signal to background ratio $S/B$ after all cuts for the $e^+e^-\mu^+\mu^-$, $2\mu^+2\mu^-$, and $2e^+2e^-$ final states at $\sqrt{s}=14$~TeV. The parameters in Eqs.~(\ref{masses}, \ref{params}) were used.}
\begin{tabular}{||l||C{0.7in}|C{0.7in}||C{0.7in}|C{0.7in}||C{0.7in}|C{0.7in}||}\hline
Channel           & \multicolumn{2}{c||}{$e^+e^-\mu^+\mu^-$} &\multicolumn{2}{c||}{$2\mu^+2\mu^-$} &\multicolumn{2}{c||}{$2e^+2e^-$} \\ \hline\hline
$\sigma$ (fb)                     & Signal  & Background  & Signal  & Background & Signal  & Background                       \\ \hline
No cuts and no energy smearing   & $0.10$ & $\cdot$        & $0.051$ & $\cdot$       & $0.051$ & $\cdot$ \\ \hline
Basic cuts~(\ref{acc.EQ}) + Trigger + Isol.~(\ref{iso.EQ})  & $0.049$ & $67$        & $0.024$ & $26$       & $0.024$ & $26$ \\ \hline
~+ $m_{4\ell}$~(\ref{Hmass.EQ}) + $m_Z^{rec}$~(\ref{Zmass.EQ}) + $m_{Z_d}^{rec}$~(\ref{Zdmass.EQ})             & $0.043$ & $0.030$     & $0.022$ & $0.017$    & $0.022$ & $0.014$ \\ \hline
$S/B$                         & \multicolumn{2}{c||}{$1.5$} &\multicolumn{2}{c||}{$1.3$} & \multicolumn{2}{c||}{$1.5$} \\ \hline
\end{tabular}
\label{tab:simulation}
\end{table*}
%%%%%%%%%%%%%%%%%%%%%%%%%%%%%%%%%%

Signal event characteristics are exploited to develop additional cuts:
\begin{itemize}
\item To resolve the four leptons of the signal, any pair of leptons is required to satisfy the isolation cut
\begin{equation}
\Delta R = \sqrt{(\Delta \eta)^2+(\Delta \phi)^2}\,>\,0.3,
\label{iso.EQ}
\end{equation}
where $\Delta\eta$ and $\Delta \phi$ are the rapidity and azimuthal angle differences between the two leptons under consideration.
\item Since signal events originate from a Higgs boson, the invariant mass of the four leptons should closely reconstruct the Higgs mass.  Using typical mass resolutions~\cite{:2012gk} for a $125$~GeV Higgs, the four lepton invariant mass, $m_{4\ell}$, is required to satisfy
\begin{equation}
|m_{4\ell} -m_H|\,<\,2\,{\rm GeV}.
\label{Hmass.EQ}
\end{equation}
\item Since the $Z$ has a relatively 
narrow width, the lepton pair identified as originating from 
the $Z$ should reconstruct $m_Z$.  Hence, we require
\begin{eqnarray}
|m_Z^{rec}-m_Z|\,<\,15\,{\rm GeV},
\label{Zmass.EQ}
\end{eqnarray}
where $m_Z^{rec}$ is the invariant mass of the reconstructed $Z$ boson.
\end{itemize}

The main backgrounds to our signal consist of the irreducible backgrounds
\begin{eqnarray}
H\rightarrow Z Z^*,\quad Z\gamma^*,\quad Z Z,\quad Z (\rightarrow 4\ell);
\end{eqnarray}
the reducible backgrounds $Zjj$, where the $Z$ decays leptonically 
and the jets fake leptons; and the $t\bar{t}$ decay chain
\begin{eqnarray}
t\bar{t}\rightarrow (b\rightarrow c \ell^-)(\bar{b}\rightarrow \bar{c}\ell^+)\ell^+\ell^-+\ptmiss,
\end{eqnarray}
where $\ptmiss$ is missing transverse momentum.
We assume that a jet fakes an electron or muon $0.1\%$ of the time~\cite{:2012gu} and use the branching ratio $\br(b\rightarrow c~\ell\bar\nu)\simeq\br(B^0\rightarrow \ell\bar\nu+{\rm anything})=0.10$~\cite{Beringer:1900zz}.  Also, the analysis of the $t\bar{t}$ background only takes into consideration the four leptons in the final state, ignoring the missing transverse momentum and extra hadronic activity from the $b$ decays.

The first row of cross sections in Table~\ref{tab:simulation} shows the signal rate before energy smearing, while in the second row the signal and background rates are given 
after energy smearing, triggers, and cuts in Eqs.~(\ref{acc.EQ},\ref{iso.EQ}).  Background cross sections before minimum cuts are not shown since soft and collinear singularities in some channels do not allow for a reliable estimate.  After these cuts, the dominant backgrounds are $t\bar{t}$ and $Z$, making up $\sim 50\%$ and $\sim 28\%$ of the $e^+e^-\mu^+\mu^-$ background and $\sim32\%$ and $\sim38\%$ of the same (lepton) flavor background, respectively.  The next largest background is $ZZ$, contributing $\sim12\%$ to $e^+e^-\mu^+\mu^-$ and $\sim 26\%$ to same (lepton) flavor backgrounds.  Unlike the other backgrounds, which require that at least one same flavor lepton pair originate from a $Z$, any flavor lepton can originate from a $b$ or $W$ in the top decay.  Hence, the different percent contributions to $e^+e^-\mu^+\mu^-$ and same flavor final states for different backgrounds are due to the combinatorics of the $t\bar{t}$ decay into four leptons.  The $t\bar{t}$, $Z$, and $ZZ$ backgrounds are mostly eliminated by the mass cuts in Eqs.~(\ref{Hmass.EQ},\ref{Zmass.EQ}), with the large majority of the leftover background coming from $Z\gamma^*$ and $H\rightarrow ZZ^*$. [See Fig.~(\ref{fig:dileptons}).]

%%%%%%%
\subsection{\boldmath $Z_d$ Resonance Peak}
%%%%%%%
In Fig.~\ref{fig:dileptons}, we show the reconstructed $Z_d$ invariant mass, $m_{Z_d}^{rec}$, distribution for major backgrounds (shown cumulatively) and our signal for the $e^+e^-\mu^+\mu^-$ signal after energy smearing, triggers, and cuts in Eqs.~(\ref{acc.EQ} - \ref{Zmass.EQ}).  The distribution is similar for the $2e^+2e^-$ and $2\mu^+2\mu^-$ signals.  As can be seen, the signal clearly stands out above background at an invariant mass of $m_{Z_d}=5$~GeV, and most backgrounds, except $Z\gamma^*$ and $H\rightarrow ZZ^*$, are negligible after the mass cuts in Eqs.~(\ref{Hmass.EQ},\ref{Zmass.EQ}).
The dilepton distributions from $H \to Z Z_d$ and $H \to Z Z^*$ in Fig.~\ref{fig:dileptons} show agreement with the results of Ref.~\cite{Davoudiasl:2012ag} which was based on Ref.~\cite{Keung:1984hn}.

The sharp fall in the invariant mass of the background below $5$~GeV can be understood by noting the invariant mass of two massless leptons can be written as
\begin{eqnarray}
m_{12}^2 = 2 E_1 E_2\,(1\,-\,\cos\theta_{12}),
\end{eqnarray}
where $E_{1,2}$ and $\theta_{12}$ are the energies and angular separation of the two leptons, respectively.  Cuts placing a minimum on the energy and the angular separation of the two leptons can be reinterpreted as a minimum on the dilepton invariant mass.  The cuts in Eqs.~(\ref{acc.EQ},\ref{iso.EQ}) effectively place a minimum on $m_{Z_d}^{rec}$, as seen in Fig.~\ref{fig:dileptons}.  Hence, to probe $Z_d$ masses below $\sim4-5$~GeV at the LHC, either the isolation or transverse momentum cuts need to be relaxed.

Motivated by Fig.~\ref{fig:dileptons}, we place the cut on the $Z_d$ reconstructed mass:
\begin{eqnarray}
|m_{Z_d}^{rec}-m_{Z_d}|\,<\,0.1\,m_{Z_d}.
\label{Zdmass.EQ}
\end{eqnarray}
To illustrate the efficiency of these cuts, Table~\ref{tab:simulation} lists the cross sections for both signal and background broken down by dilepton final state signals.   The second row shows the cross sections after energy smearing, triggers, and cuts in Eqs.~(\ref{acc.EQ},\ref{iso.EQ}), while the third row includes the additional invariant mass cuts of Eqs.~(\ref{Hmass.EQ},\ref{Zmass.EQ},\ref{Zdmass.EQ}).  In the last row we show the signal to background ratio, $S/B$, after all cuts.   As can be clearly seen, the invariant mass cuts leave the signal cross section mostly intact while severely suppressing the backgrounds, and the $S/B$ ratio is well under control.

%%%%%%%%%%%%% TABLE %%%%%%%%%%%%%
\begin{table}[b]
\centering
\caption{Luminosities needed for $2\sigma$ exclusion, $3\sigma$ observation, and $5\sigma$ discovery after all cuts for both a $m_{Z_d}=5$~GeV and $m_{Z_d}=10$~GeV at $\sqrt{s}=14$~TeV with and without $K$-factors.  All other parameters are the same as in Eqs.~(\ref{masses}, \ref{params}).}
\begin{tabular}{|c|C{.8in}|C{.8in}|C{.8in}||}\hline
&\multicolumn{3}{|c||}{$m_{Z_d}=5$~GeV} \\\hline
 & 2$\sigma$ (Excl.) & 3$\sigma$ (Obs.) & 5$\sigma$ (Disc.)  \\ \hline
No $K$-factors& 78 fb$^{-1}$ & 180 fb$^{-1}$ & 490 fb$^{-1}$\\ \hline
+$K$-factors & 33 fb$^{-1}$ & 75 fb$^{-1}$ & 210 fb$^{-1}$\\\hline  \hline
&\multicolumn{3}{|c||}{$m_{Z_d}=10$~GeV} \\\hline
&2$\sigma$ (Excl.) & 3$\sigma$ (Obs.) & 5$\sigma$ (Disc.)  \\ \hline
No $K$-factors&100 fb$^{-1}$ &230 fb$^{-1}$ & 640 fb$^{-1}$ \\ \hline
+$K$-factors & 42 fb$^{-1}$ &95 fb$^{-1}$ & 260 fb$^{-1}$ \\ \hline
\end{tabular}
\label{tab:signif}
\end{table}
%%%%%%%%%%%%%%%%%%%%%%%%%%%%%%%%%
%%%%%%%%%%%%% FIGURE %%%%%%%%%%%%%
\begin{figure}[t]
\begin{center}
\includegraphics[width=0.45\textwidth,clip]{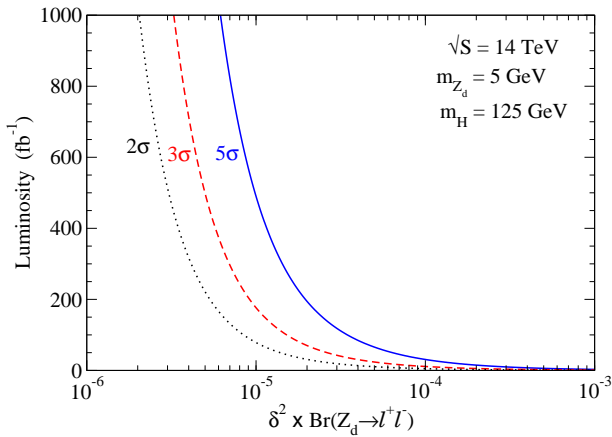}
\end{center}
\caption{Luminosities needed for a $2\sigma$ exclusion (dotted), 3$\sigma$ observation (dashed), and 5$\sigma$ discovery (solid) as a function of $\delta^2\times \br(Z_d \to \ell^+ \ell^-)$ at $\sqrt{s}=14$~TeV.  All other parameters are the same as in Eqs.~(\ref{masses}, \ref{params}).}
\label{fig:signif}
\end{figure}
%%%%%%%%%%%%%%%%%%%%%%%%%%%%%%%%%%

We also explore how sensitive the LHC is to the parameters $m_{Z_d}$ and $\delta^2\times \br(Z_d \to \ell^+ \ell^-)$.  In Table~\ref{tab:signif} we illustrate the luminosities needed at $\sqrt{s}=14$~TeV for 2$\sigma$ exclusion, 3$\sigma$ observation, and 5$\sigma$ discovery for both $m_{Z_d}=5$~GeV and $m_{Z_d}=10$~GeV. Significances have been calculated as $S/\sqrt{S+B}$.  All other parameters are the same as in Eqs.~(\ref{masses}, \ref{params}).  For a slightly heavier $Z_d$ the LHC is found to be less sensitive.   The decrease in sensitivity with increasing mass can be understood in part by noting that for a higher $m_{Z_d}$ the cut in Eq.~(\ref{Zdmass.EQ}) becomes less stringent.  For our parameterization the signal rate is the same for both $Z_d$ masses.  Since the cumulative background is relatively flat in the region of interest,  the amount surviving cuts increases.  Hence, the significance slightly decreases as $m_{Z_d}$ increases.  Note, however, that the same $\delta^2\times \br(Z_d \to \ell^+ \ell^-)=10^{-5}$ 
was used for both $Z_d$ masses.  It is quite plausible, perhaps even likely, that $\delta$ is proportional to $\mzd$.  
In that case, the 10~GeV signal will be enhanced by a factor of 4, requiring $6-7$ times less than the luminosities listed in Table~\ref{tab:signif}.

In Fig.~\ref{fig:signif}, the luminosity needed for a 2$\sigma$ exclusion (dotted), 3$\sigma$ observation (dashed), and 5$\sigma$ discovery (solid) is plotted as a function of $\delta^2\times \br(Z_d \to \ell^+ \ell^-)$ keeping $m_{Z_d}=5$~GeV and all other parameters the same as Eqs.~(\ref{masses}, \ref{params}).  The sensitivity of the LHC dramatically decreases as $\delta^2\times \br(Z_d \to \ell^+ \ell^-)$ decreases.

The effects considered so far have been at leading order.  However, including higher order QCD corrections can increase our sensitivity at the LHC.  To approximate these corrections, some standard $K$-factors~\footnote{$K \sim 2$ for $g g \to H$ at NNLO-NNLL in $\alpha_s$~ \cite{Harlander:2002wh,Anastasiou:2002yz} (N=Next-to; LO=Leading Order; LL=Leading Log), $K = 1.4$ for $t \bar t$ at NNLO-NNLL \cite{Czakon:2013goa}, and $K=1.19$ for $Z$ at NNLO~\cite{Harlander:2002wh,Hamberg:1990np}. $K=1.18$ for $Z\gamma$ at NLO in $\alpha_s$~\cite{Campbell:2011bn,Ohnemus:1992jn}, $K=1.62$ for $ZZ$ at NLO~\cite{Campbell:2011bn,Ohnemus:1990za}, $K=0.9$ for $Zjj$ at NLO~\cite{Campbell:2003hd}.} are applied.  The luminosities needed for a 2$\sigma$ exclusion, 3$\sigma$ observation, and 5$\sigma$ discovery after the inclusion of these $K$-factors are shown in     Table~\ref{tab:signif}.  As can be seen, the inclusion of higher order corrections can greatly increase the sensitivity of the LHC to these processes, decreasing the luminosity needed for an observation by roughly $60\%$.

From Table~\ref{tab:signif} and Fig.~\ref{fig:signif}, we conclude that with 
$\sim~{\rm few} \times 100$ fb$^{-1}$ the LHC is sensitive to the four lepton process $H\rightarrow ZZ_d$ for $Z_d$ masses in the range $5-10$~GeV and can probe $\delta^2\times \br(Z_d \to \ell^+ \ell^-)$ down to about $10^{-5}$.  The rescaling of these limits for nonzero $\kappa_X$ and $\tilde{\kappa}_X$ needs to take into account the effects of the cuts.  As discussed in the next section, the transverse momentum cuts in Eq.~(\ref{acc.EQ}) cut more signal from operators of class (B) than class (A), due to the transverse polarization of the $Z_d$'s.  The cuts of Eqs.~(\ref{acc.EQ},\ref{iso.EQ}) and triggers decrease the signal rate of operators of class (A) by $\sim 50\%$ [see Table~\ref{tab:simulation}] and operators of class (B) by $\sim 65\%$.  These limits can also simply be scaled to generalize to the case with different branching ratios of $Z_d$ into leptons.

%%%%%%%
\subsection{Angular Distribution}
%%%%%%%
%%%%%%%%%%%%%%%%%%%%% FIGURE %%%%%%%%%%%%%
\begin{figure*}[tb]
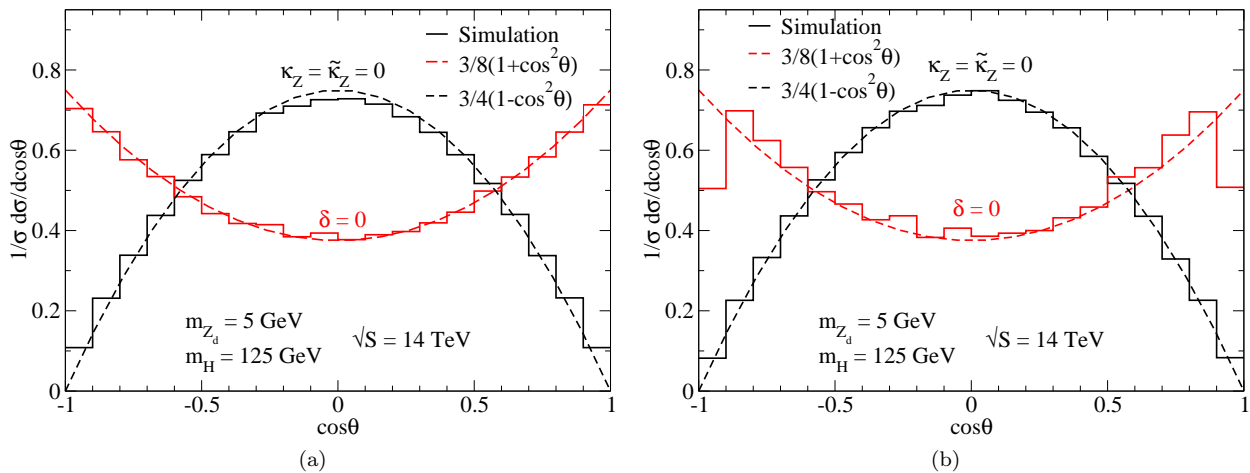

\begin{center}
\subfigure[]{
      \includegraphics[width=0.45\textwidth,clip]{cthlNCNS.eps}
}
\subfigure[]{
      \includegraphics[width=0.45\textwidth,clip]{cthl_all.eps}
}
\caption{Solid histograms are event simulation results for angular distributions of the $\ell^-$ identified as originating from the $Z$ in the reconstructed $Z$ rest frame (a) without cuts or smearing and (b) with energy smearing, triggers, and cuts from Eqs.~(\ref{acc.EQ} - \ref{Zdmass.EQ}).  The angle is measured with respect to the $Z$ moving direction in the partonic c.m. frame.  Dashed lines are expected distribution for longitudinally (black) and transversely (red/grey) polarized $Z_d$'s.}
\label{fig:angdist}
\end{center}
\end{figure*}
%%%%%%%%%%%%%%%%%%%%%%%%%%%%%%%%%%

As mentioned earlier, interactions mediated by the operators of class (A) are expected to be dominated by the longitudinal polarizations of the $Z_d$, while the operators of class (B) are dominated by its transverse polarizations.  Once a $H\rightarrow Z Z_d$ signal is discovered, determining the polarization of the $Z_d$ can allow us to discriminate whether the interaction is originating from the operators of class (A) or (B).

A good diagnostic of vector boson polarization is the angular distribution of its decay products.  In the $Z_d$ rest frame, we have
\begin{eqnarray}
\frac{d\Gamma(Z_d\rightarrow\ell^+\ell^-)}{d\cos\theta_\ell}\sim (1\pm\cos^2\theta_\ell)
\label{angdist.EQ}
\end{eqnarray}
where $\theta_\ell$ is the lepton ($\ell^-$) angle with respect to the $Z_d$ spin-quantization axis, and the upper (lower) sign is for a transversely (longitudinally) polarized $Z_d$.  Events produced via operators of class (A) and class (B) are therefore expected to result in different leptonic angular distributions of the $Z_d$ decay products, allowing for a distinction between the two cases.  To exploit this effect, a spin-quantization axis must be chosen such that the $Z_d$ is mostly longitudinally or transversely polarized with respect to that axis.

For $m_{Z_d} \ll m_H - m_Z$, the $Z_d$ will be produced highly boosted and is well approximated as a helicity eigenstate.  That is, the $Z_d$ is mostly in a longitudinally or transversely polarized state with respect to its moving direction.  One might expect to be able to use the $Z_d$ moving direction as the spin-quantization axis and use the angular distribution of Eq.~(\ref{angdist.EQ}) to measure the $Z_d$ polarization.  However, when $\cos\theta_\ell\sim\pm1$, one lepton is moving with the direction of motion of the $Z_d$ in the lab frame and the other lepton in the opposite direction.  The boost from the $Z_d$ rest frame to the lab frame is then against the direction of motion of one of the leptons.  Hence, this configuration results in events with the softest leptons and the transverse momentum cut of Eq.~(\ref{acc.EQ}) eliminates the events with $\cos\theta_\ell\sim\pm1$.  Since this region is vital in distinguishing between our two cases, it will be useful to use another angular distribution directly related to that of Eq.~(\ref{angdist.EQ}).

In the partonic center of momentum (c.m.) frame, the $Z$ and $Z_d$ momentum and spin must be anti-aligned  by conservation of momentum and angular momentum, respectively.  As a result, if the $Z_d$ is in a helicity eigenstate, then in the partonic c.m. the $Z$ is also in the same helicity eigenstate.  The angular distribution of $Z$ decay products will be of the same form as Eq.~(\ref{angdist.EQ}), now with the angle measured in the $Z$ rest frame with respect to the $Z$ moving direction in the partonic c.m. frame.   Since the decay products of the $Z$ are typically harder than those of the $Z_d$, the $p_T$ cuts are not severe and the angular distribution is a good diagnostic of the $Z_d$ polarization.  

In Fig.~\ref{fig:angdist}, we show the simulated angular distribution of the lepton identified as originating from the $Z$ decay for operators of class (A) (solid black) and class (B) (solid red/grey), and the expected distributions for longitudinally and transversely polarized $Z_d$ (dashed black and red/grey, respectively).  Both operators of class (B), Eqs.~(\ref{caseB},\ref{caseBtil}), result in the same angular distributions as we take either $\kappa_Z \ne 0$, $\tilde \kappa_Z = 0$ or $\kappa_Z = 0$, $\tilde \kappa_Z \ne 0$.
The CP violating effect would show up as an interference effect when both operators are present.  Note that for $\delta =0$, the value of $\kappa_Z$ and $\tilde \kappa_Z$ will effect the total rate of our process but not the distributions in Fig.~\ref{fig:angdist}, since they have been normalized.

These angular distributions are measured in the reconstructed $Z$ rest frame with respect to the $Z$ moving direction in the partonic c.m. frame.  As can be seen in Fig.~\ref{fig:angdist} (a), without energy smearing or cuts it is clear that, using the above definition, the angular distribution of the $Z$ decay products reflect the expected $Z_d$ polarization.  That is, operators of class (A) result in events that are dominated by longitudinal $Z$'s and $Z_d$'s, while the operators of class (B) are dominated by transversely polarized $Z$'s and $Z_d$'s.  

Figure~\ref{fig:angdist} (b) illustrates  the effects of the cuts on the angular distributions.  Even after cuts, the two distributions are still clearly distinguishable.  The small relative depletion of events at $\cos\theta_\ell\sim\pm1$ is due to the isolation cuts in Eq.~(\ref{iso.EQ}).  This can be understood by noting that for $\cos\theta_\ell\sim\pm1$ in the $Z$ rest frame, one lepton from the $Z$ decay is moving directly against the $Z$ moving direction in the partonic c.m. frame.  Also, since $m_Z\sim m_H$, the $Z$ is produced nearly at threshold and its decay products move back-to-back in the partonic c.m. frame. Hence,  for $\cos\theta_\ell\sim\pm1$, one of the leptons from $Z$ decay is typically moving in the $Z_d$ direction.  In this configuration, $\Delta R$ between the $Z$ and $Z_d$ decay products is minimized and fail the cut in Eq.~(\ref{iso.EQ}).

%%%%%%%%%%%%%%%%%%%%%%%%%%%%
\section{Conclusions}
%%%%%%%%%%%%%%%%%%%%%%%%%%%%
Recent data from the LHC appears to have uncovered a Higgs scalar associated with the mechanism responsible for spontaneous electroweak symmetry breaking.  The new scalar $H$ may exhibit properties that lead to deviations from the Standard Model (SM) predictions. Early findings suggest inconclusive yet tantalizing hints for such deviations 
in Higgs decays.  Regardless of the fate of these hints, searching for physics beyond the SM is well motivated, especially in light of the need to account for the cosmic dark matter density in the Universe.  It is reasonable to expect that the ``dark" or ``hidden"  sector of particle physics, like its visible counter part, is endowed with structure and its own forces.  In fact, certain astrophysical observations have been interpreted as signals of dark matter that couples to a hidden sector light vector boson. One can then ask whether the forces in the dark sector could manifest themselves through their interactions with the Higgs.

Based on the above motivation, in this work we have studied the possibility that a dark vector boson $Z_d$, in the mass range of $5-10 ~\gev$, could couple to $H$ via mixing or through loop effects.  These couplings can then be described by two classes of operators, leading to Higgs decays into dominantly longitudinal or transverse $Z_d$.  An interesting typical decay in the first class is $H\to Z Z_d$, while the second class of decays could include $H\to X Z_d$, with $X=Z, Z_d, \gamma$.

We focused on $H\to Z Z_d$, as a representative novel decay channel.  Using leptonic final states for both $Z$ and $Z_d$, we found that the next run of the LHC is capable of excluding or detecting such decays for $\delta^2\times \br(Z_d \rightarrow \ell^+\ell^-)\sim10^{-5}$ and $m_{Z_d}\sim 5-10$~GeV or loop induced dimension 5 operators of similar magnitude, with a few hundred fb$^{-1}$ of data.  The branching ratios of $Z_d$ in leptons can be as large as $\br(Z_d \rightarrow e^+e^-) \simeq \br(Z_d \rightarrow \mu^+\mu^-) \simeq 0.15$.   For somewhat lower $m_{Z_d}$, larger backgrounds will probably require longer running.  These LHC searches via rare Higgs decays provide a complementary approach to low energy experiments that are primarily sensitive to dark vector boson masses at or below the GeV scale.

Our results also suggest that one could use the angular distribution of the leptons from the $Z$ in $H\to Z Z_d$ to probe the underlying micro-physical interaction.  Those distinct distributions can be used to determine whether the $Z_d$ emitted in the decay was longitudinal or transverse, providing a probe of the nature of the Higgs coupling to the dark sector. 

In the various scenarios we have considered, the Higgs boson can also decay into $\gamma Z_d$ and $Z_d Z_d$.  Although we have not discussed those modes in detail here, they
entail very distinct signatures \cite{Davoudiasl:2012ig} that should help separate them from background and ordinary SM Higgs decays. For example, in the case of very light final state $Z_d$ bosons, they look like promptly converted $e^+ e^-$ photons; a feature that distinguishes them from ordinary diphoton events.  Heavier $Z_d$ decays could stand out above Dalitz and  four lepton ordinary decays of the Higgs, depending on their abundance in the data.

We hope that our work encourages further examination of the Higgs properties at the LHC, or future facilities, as a potential means of peering into the hidden sector and shedding light on the nature of dark matter or other as yet unknown phenomena.

\FloatBarrier

%---------------------------------------------------------
Acknowledgments:
%---------------------------------------------------------
We would like to thank Jay Hubisz for useful discussions and Bogdan Wojtsekhowski for useful comments.  This work was supported in part by the United States DOE under Grant No.~DE-AC02-98CH10886, No.~DE-AC05-06OR23177, and by the NSF under Grant No.~PHY-1068008.
WM acknowledges partial support as a Fellow in the Gutenberg Research College.

%%%%%%%%%%%%%%%%%%%%%%%%%%%%%%%%%%%%%%%%%%%%%%%%%%%%%%%%
\begin{widetext}
\appendix*
%%%%%%%%%%%%%%%%%%%%%%%%%%%%%%%%%%%%%%%%%%%%%%%%%%%%%%%%
\section{Various relevant Higgs decay widths}

%%%%%%%
\subsection{\boldmath $H \to Z\, Z_d$ decay}
%%%%%%%
The amplitude for $H\to Z(p)Z_d(q)$ is
\beq
i {\cal M} = i {\cal C}_{HZZ}^\text{SM} K_{ZZ_d,\mu\nu} \epsilon^{*\nu} (p)\epsilon^{*\mu} (q)
\eeq
where ${\cal C}_{HZZ}^\text{SM} \equiv g m_Z / \cos\theta_W$ is the SM $HZZ$ coupling and 
\beq
K_{ZZ_d,\mu\nu} = \left( \eps_Z g_{\mu\nu} + \frac{\kappa}{m_Z^2} \left(p \cdot q g_{\mu\nu} - p_\mu q_\nu\right) + \frac{\tilde\kappa}{m_Z^2} \eps_{\mu\nu\rho\sigma} p^\rho q^\sigma \right). 
\label{effFeynmanRule}
\eeq
Here, tree-level mixing ($\eps_Z$) is from dimension 3 operators [Eq.~(\ref{kappaA})] and loop-induced couplings 
($\kappa_Z$ for CP conserving part and $\tilde\kappa_Z$ 
for CP violating part) is from dimension 5 operators [Eqs.~(\ref{kappaB}, \ref{kappaBtilde})].

We obtain, after summing over polarizations
\bea
\sum_{\rm pol}|{\cal M}|^2 &=& \left({\cal C}_{HZZ}^\text{SM}\right)^2\\ 
&\times&\left[ \eps_Z^2 \left(\frac{(p \cdot q)^2}{m_Z^2 m_{Z_d}^2} + 2\right) + 6 \eps_Z \frac{\kappa_Z}{m_Z^2} (p \cdot q) +   \left(\frac{\kappa_Z}{m_Z^2}\right)^2 \left(m_Z^2 m_{Z_d}^2 + 2 (p \cdot q)^2\right) - 2 \left(\frac{\tilde\kappa_Z}{m_Z^2}\right)^2 \left( m_Z^2 m_{Z_d}^2 - (p \cdot q)^2 \right) \right]\nonumber
\eea
with $p \cdot q = \frac{1}{2} \left(m_H^2 - m_Z^2 - m_{Z_d}^2\right)$ for $H \to Z Z_d$ decay.

For $m_{Z_d}^2 \ll D^2$ (with $D^2 \equiv m_H^2 - m_Z^2$), we have
\bea
\Gamma(H \to Z Z_d) &=& 4 \pi \frac{\sqrt{\lambda(m_H^2, m_Z^2, m_{Z_d}^2)}}{64 \pi^2 m_H^3} \,\sum_{\rm pol}|{\cal M}|^2 \\
&\simeq& \frac{D^2}{16 \pi m_H^3} \left({\cal C}_{HZZ}^\text{SM}\right)^2 \left( \eps_Z^2 \frac{D^4}{4 m_Z^2 m_{Z_d}^2} + 3 \eps_Z \kappa_Z \frac{D^2}{m_Z^2} + \kappa_Z^2 \frac{D^4}{2 m_Z^4} + \tilde \kappa_Z^2 \frac{D^4}{2 m_Z^4} \right)
\eea
with $\lambda(x,y,z) = x^2 + y^2 + z^2 - 2xy - 2yz - 2zx$.

As expected, the longitudinal polarization limit ($\kappa_Z, \tilde\kappa_Z \to 0$) shows an 
enhancement as $m_{Z_d} \to 0$, in accordance with 
the Goldstone Boson Equivalence Theorem, 
while the transverse polarization limit ($\eps_Z \to 0$) does not.  However, because $\eps_Z = \delta\, \mzd/m_Z$, 
the rate is not singular.

%%%%%%%
\subsection{\boldmath $H \to Z_d\, Z_d$ decay}
%%%%%%%
The amplitude for the decay $H \to Z_d(p)Z_d(q)$ is given by 
\beq
i {\cal M} = i {\cal C}_{HZZ}^\text{SM} K_{Z_dZ_d,\mu\nu} \epsilon^{*\nu} (p)\epsilon^{*\mu} (q).
\eeq
When operators of class (A) arise from $Z$-$Z_d$ mass mixing, two insertions of the mixing angle are needed to obtain a $H$-$Z_d$-$Z_d$ coupling.  This is equivalent to the replacement $\eps_Z\to \eps^2_Z$.  Since operators of class (B) arise from loops, the $H$-$Z_d$-$Z_d$ and $H$-$Z$-$Z_d$ interactions are of the same form in this case.    The effective Feynman rule for the $H$-$Z_d$-$Z_d$ coupling is of the same form as the right-hand side of Eq.~\eqref{effFeynmanRule} with $\eps_Z\to\eps^2_Z$ and we get
\beq
\sum_{\rm pol}|{\cal M}|^2 = \left({\cal C}_{HZZ}^\text{SM}\right)^2 \left[ \eps_Z^4 \left(\frac{(p \cdot q)^2}{m_{Z_d}^4} + 2\right) + 6 \eps_Z^2 \frac{\kappa_{Z_d}}{m_Z^2} (p \cdot q) +   \left(\frac{\kappa_{Z_d}}{m_Z^2}\right)^2 \left(m_{Z_d}^4 + 2 (p \cdot q)^2\right) - 2 \left(\frac{\tilde\kappa_{Z_d}}{m_Z^2}\right)^2 \left( m_{Z_d}^4 - (p \cdot q)^2 \right) \right]
\eeq
with $p \cdot q = \frac{1}{2} \left(m_H^2 - 2 m_{Z_d}^2\right)$ for $H \to Z_d Z_d$ decay.  For the case of scalar mixing in the Higgs sector see Ref.~\cite{Gopalakrishna:2008dv}.

For $m_{Z_d}^2 \ll D^2$ (with $D^2 \equiv m_H^2$), we have 
\bea
\Gamma(H \to Z_d Z_d) &=& \frac{4 \pi}{2} \frac{\sqrt{\lambda(m_H^2, m_{Z_d}^2, m_{Z_d}^2)}}{64 \pi^2 m_H^3} \,\sum_{\rm pol}|{\cal M}|^2 \\
&\simeq& \frac{D^2}{32 \pi m_H^3} \left({\cal C}_{HZZ}^\text{SM}\right)^2 \left( \eps_Z^4 \frac{D^4}{4 m_{Z_d}^4} + 3 \eps_Z^2 \kappa_{Z_d} \frac{D^2}{m_Z^2} + \kappa_{Z_d}^2 \frac{D^4}{2 m_Z^4} + \tilde \kappa_{Z_d}^2 \frac{D^4}{2 m_Z^4} \right) .
\eea

%%%%%%%
\subsection{\boldmath $H \to \gamma\, Z_d$ decay}
%%%%%%%
The amplitude for $H\to \gamma(p) Z_d(q)$ is given by 
\beq
i {\cal M} = i {\cal C}_{HZZ}^\text{SM} K_{\gamma Z_d,\mu\nu} \epsilon^{*\nu} (p)\epsilon^{*\mu} (q)
\eeq
with the effective Feynman rule of $H$-$\gamma$-$Z_d$ coupling being the same form as in 
the right-hand side of Eq.~\eqref{effFeynmanRule} except that there is is no $\eps_Z$ term here.

We then have 
\bea
\sum_{\rm pol}|{\cal M}|^2 &=&2 \left({\cal C}_{HZZ}^\text{SM}\right)^2 \left(\kappa_\gamma^2  
+ \tilde\kappa_\gamma^2  \right)\left(\frac{p\cdot q}{m_Z^2}\right)^2
\eea
with $p \cdot q = \frac{1}{2} \left(m_H^2 - m_{Z_d}^2\right)$ for $H \to \gamma Z_d$ decay.

For $m_{Z_d}^2 \ll D^2$ (with $D^2 \equiv m_H^2$), we obtain
\bea
\Gamma(H \to \gamma Z_d) &=& 4 \pi \frac{\sqrt{\lambda(m_H^2, 0, m_{Z_d}^2)}}{64 \pi^2 m_H^3} \,\sum_{\rm pol}|{\cal M}|^2 \\
&\simeq& \frac{1}{32 \pi} \left({\cal C}_{HZZ}^\text{SM}\right)^2 \frac{D^6}{m_H^3 m_Z^4} 
\left( \kappa_\gamma^2 + \tilde\kappa_\gamma^2 \right) .
\eea

As expected, there is no enhancement from the longitudinal polarization of $Z_d$ in this case.

\end{widetext}

%---------------------------------------------------------

%================================
% END THE DOCUMENT
%================================

\end{document}